\documentclass[structabstract]{aa}  
\usepackage{graphicx}
\usepackage{txfonts}
\usepackage{natbib}
\bibpunct{(}{)}{;}{a}{}{,} 
\newcommand{\bd}{BD+33$^{\circ}$2642}

\begin{document}
\title{Binary central stars of planetary nebulae with long orbits: 
the radial velocity orbit of
BD+33$^{\circ}$2642  (PN G052.7+50.7) and the orbital motion of
HD\,112313 (PN LoTr5). \thanks{Based on observations made with the Mercator
  Telescope, operated on the island of La Palma by the Flemish
  Community, at the Spanish Observatorio del Roque de los Muchachos of
  the Instituto de Astrof\'isica de Canarias.}}
\titlerunning{Binary PNe}
\author{Hans Van Winckel\inst{1}, Alain Jorissen\inst{2}, Katrina
  Exter\inst{1}, Gert Raskin\inst{1}, Saskia Prins\inst{1}, Jesus
  Perez Padilla \inst{1}, Florian Merges\inst{1}, Wim Pessemier\inst{1}}
\institute{Instituut voor Sterrenkunde, KU Leuven, Celestijnenlaan
  200D bus 2401, B-3001 Leuven, Belgium \and Institut d'Astronomie et
  d'Astrophysique, Universit\'{e} Libre de Bruxelles,  CP 226,
  Boulevard du Triomphe, 1050, Bruxelles, Belgium}

\offprints{Hans.VanWinckel@ster.kuleuven.be}
\date{}
\authorrunning{H. Van Winckel et al.}
\titlerunning{Spectroscopic orbits of \bd\, and HD\,112313}

\abstract {} {We study the impact of binary interaction
  processes on the evolution of low- and intermediate-mass stars using
  long-term monitoring of their radial velocity. Here
  we report on our results on the central stars of two planetary
  nebulae (PNe): the well-studied spectrophotometric standard
  BD+33$^{\circ}$2642 (central star of PNG 052.7+50.7) and
    HD\,112313 (central star of PN LoTr5), the optical light of which is
  dominated by a rapidly rotating G star.}  
{The high-resolution
  spectra were cross-correlated with carefully selected masks of
  spectral lines. The individual masks were optimised for the spectral
  signatures of the dominant contributor of the optical light.}  
{We report on the first detection of orbital motion in these two
  objects.
  For \bd\,  we sampled 1.5 cycles of the 1105 $\pm$ 24 day 
  orbital period. For HD\,112313 a full period is not yet covered, despite
  our 1807 days of monitoring. The radial-velocity
  amplitude shows that it is unlikely that the orbital plane is
  co-planar with the one defined by the nebular waist of the bipolar
  nebula. To our knowledge these are the first
  detections of orbits in PNe that are in a range from several weeks to a few
  years.} 
{The orbital
  properties and chemical composition of \bd\, are similar to what is found in post-AGB binaries with
  circumbinary discs. The latter are probably progenitors of
  these PNe. For LoTr5 the Ba-rich central star and the long
  orbital period are similar to the Ba star giants, which hence serve as
natural progeny. In contrast to the central star in LoTr5, normal Ba
stars are slow rotators. The orbits of these systems have a low
probability of occurrence according to recent population synthesis calculations. }


\maketitle

\section{Introduction}
The role of binary interaction processes in the shape and shaping of
PNe and their progenitors, the proto-planetary nebulae (PPNe), is far
from understood \citep[e.g.][and references
therein]{balick02,vanwinckel03,demarco09}.  There is growing consensus
that binary interaction processes play a fundamental role in many
objects, but the theoretical understanding is subject to many
unsupported assumptions such as the efficiency of envelope ejection,
the physical description of the common-envelope phase (CE phase,
\citet{izzard12,ivanova13}), the accretion efficiency onto the
companion \citep{ricker08}, and the jet formation mechanisms in
binaries. It also is an observational challenge to constrain the
requirements for the most important shaping agents such as jets or
magnetic fields and determine whether they are active or not.

Direct observational evidence of the binary nature of the central star
of PNe and obscured PPNe is notoriously difficult to obtain
\citep{demarco09} and all the techniques have their specific biases.
The most successful method so far is the detection of orbital
modulation in the light curve
\citep[e.g.][]{bond00b,demarco08,miszalski09a,miszalski09b,miszalski11b},
which led to the discovery of about 40 systems. The modulation can be
caused by ellipsoidal deformation, eclipses, or irradiation of the
cool component by the hot component \citep[e.g.][]{hillwig10}. These
photometric methods are only sensitive to systems that have
spiraled-in during a CE phase and have now a period shorter than about
10 days. Several of these post-CE systems also show jets powered by
accretion \citep[e.g.][]{boffin12, miszalski13a}.  It is as yet
unclear under which conditions wider systems will create
accretion-powered jets; the main reason is the lack of
observational data on the orbital characteristics of the central
stars.

Here we report on the first discovery of two wide spectroscopic
binaries among central stars of PNe. We introduce the objects
(Sect.~\ref{sect:objects}) and our radial-velocity programme
(Sect~\ref{sect:radvel}) in the next sections. We focus on the results
in Sect.~\ref{sect:bd} and Sect.~\ref{sect:LoTr5} and
conclude in Sect.~\ref{sect:results}.

\section{Programme stars} \label{sect:objects}

\subsection{\bd\, (PNG 052.7+50.7) }
\bd\,  is the central star of one of the rare halo PNe (PNG
052.7+50.7). It is very well observed because it
is a spectrophotometric standard star \citep{oke90}. The roundish faint 
nebula of about
5\arcsec\, in diameter was discovered by \cite{napiwotzki93} and found to be
consistent with the characteristics of the central star with a
T$_{\rm eff}$ of 20200\,K. Despite this rather low temperature, this
star is thought to be the ionising source of the PN.
Interestingly, the central star is now very metal
poor, but the initial metallicity was most likely higher: the object shows
a depletion of the refractory elements \citep{napiwotzki94} similar to
what is found in many post-AGB binaries \citep[e.g.][and references
therein]{vanwinckel03,giridhar05,vanwinckel09}. The abundances can be
explained by assuming an accretion phase of circumstellar gas that was
subject to a phase of dust production. The refractory elements are
preferentially locked in the dust grains, and only the gas cleaned
from the dust is accreted onto the star. 
For post-AGB stars, this depletion pattern is
observed in binaries with circumbinary discs \citep{vanwinckel95,
  vanwinckel03}.

The object has been subject to radial-velocity monitoring
\citep{demarco04, demarco07} but the data were too scarce to draw firm
conclusions on the possible binary nature of the object. Between
August 2002 and September 2003, 14 measurements were reported with a
peak-to-peak amplitude of 15.3 km\,s$^{-1}$. In 2005, radial velocity
variability was detected on a time scale of about four days and with an
amplitude of $\sim$ 5 km\,s$^{-1}$, but this variability was absent
from the short data set obtained in 2006. A clear offset of 10
km\,s$^{-1}$ was detected between the mean velocities of the two
different runs in 2005 and 2006, which were about 420 days apart.

\subsection{HD\,112313 (PN LoTr5)}
HD\,112313 (BD+26$^{\circ}$2405, central star of PN LoTr5 ) was
discovered by \cite{longmore80} and is known to be a binary PN at high
Galactic latitude because the optical light is dominated by a rapidly
rotating G5 III star \citep[e.g.][]{bond03}. The source of the PN is a
very hot central white dwarf (WD; \cite{feibelman83}). The rapidly
rotating cold component ($v\,\sin\,i \sim$\,65 km\,s$^{-1}$) is
enriched in s-process elements \citep{thevenin97} and hence can be
classified as a Ba star. The observed 5.9-day period in the photometry
is consistent with the rotation of the giant \citep{jasniewicz97,
  strassmeier97}. The system has a long history of period
determinations with earlier claims that the cool component was a
narrow system \cite[e.g.][and references therein]{strassmeier97}, but
so far without compelling evidence for either a short binary period, or
for the claimed triple nature of the system.  The X-ray spectrum is
too hard to be emitted by the WD and probably originates in the
coronal activity of the rapidly rotating G5 giant of the system
\citep{montez10}.  Long-slit spectra of the faint large nebula were
modelled successfully by \cite{graham04}, who assumed a bipolar outflow
with the polar axes inclined by only 17$^{\circ}$ to the line of sight
(see their Fig.~7).

\section{Radial-velocity programme} \label{sect:radvel}
We used the HERMES spectrograph \citep{raskin11} mounted on the 1.2m
Mercator telescope. This highly efficient, fibre-fed spectrograph is
located in a temperature-controlled environment. 
We used in this programme the high-resolution science fibre,
yielding a spectral resolution ($\lambda/\Delta\,\lambda$) of 85000
over the whole wavelength domain from 377 to 900 nm.  We began operating
our spectrograph in June 2009 and since then have accumulated data in our
large radial-velocity monitoring programme, which covers a wide range of
evolved objects \citep{vanwinckel10}. The zero-point of our
radial-velocity determinations are scaled to the IAU radial-velocity
scale \citep[e.g.][]{chubak12}. Our instrumental zero-point is
subject to pressure variations during the night \citep{raskin11} and
we obtain a standard deviation of 70 m\,s$^{-1}$\, based on 2137
measurements of different radial-velocity standards
and measured as a spread over the mean of every standard. These data
cover the period from mid-2009 onward to August 2013.

In this Letter we focus on the results obtained on two central stars
of PNe, and the characteristics of the two data sets are given in
Table~\ref{tab:data}. The requirements on both the length and the
sampling of this programme is a non-trivial observational
challenge. With our own 1.2m Mercator telescope we monitor the targets
throughout the year and can adjust the sampling rate to the expected
variability \footnote{The radial velocity data is available in
    electronic form at the CDS via {\texttt{http://cdsweb.u-strasbg.fr/cgi-bin/qcat?J/A+A/}}}.

\begin{table}
\caption{Overview of the monitoring data for the two objects.}
\centering
\label{tab:data}                    
\begin{tabular}{lrll}        
\hline\hline                 
Central Star   & \# & JD first & time span (d) \\ \hline
\bd   & 202 & 2455014.43 & 1684  \\
HD\,112313 & 50  & 2454885.72 & 1807  \\
\hline \hline
\end{tabular}
\end{table}

\section{Analysis of BD+33$^{\circ}$2642} \label{sect:bd}

\bd\, is a relatively bright central star ($V$ = 10.7 mag). Our radial
velocity was computed using a discrete cross-correlation technique in
which the input is a list of spectral lines. The method does not include
an a priori model of the spectral broadening function. The routine works in
pixel-order space and on the basis of the extracted spectrum. The
contribution of every line to the final cross-correlation function is 
weighted by the local signal.

For \bd\, we optimised a spectral mask using the lines identified by
\cite{napiwotzki94} in their chemical analysis of the central star.
We omitted the lines that are affected by the nebular component. The
cross-correlation function is well defined and an individual
measurement has an average intrinsic error of 270 m\,s$^{-1}$. 

\begin{figure}
\begin{center}
\resizebox{.70\hsize}{!}{ \includegraphics{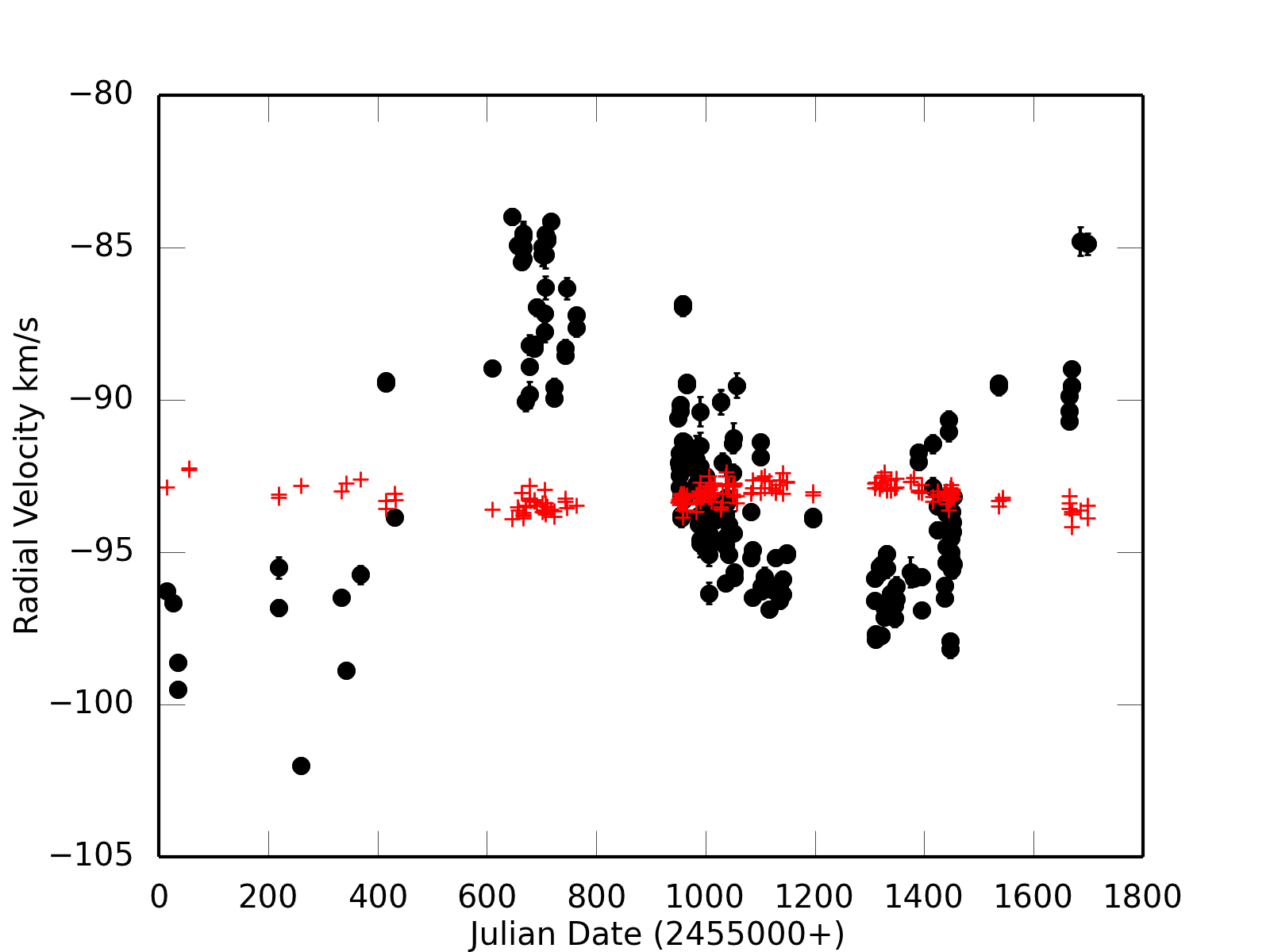}}
\caption{Time series of barycentric radial velocity measurements
  of \bd. The crosses indicate the nebular velocity determined using the
  H$\alpha$ emission line.}
\label{fig:BDTime}
\end{center}
\end{figure}

In Fig.~\ref{fig:BDTime} we show the full time series of our radial-velocity programme. The crosses in the figure indicate the radial velocity of the nebular
H$\alpha$ emission line with a rest wavelength of
$\lambda$6562.8\,\AA. There is a clear long-term trend in the
photospheric data. We interpret
this as caused by orbital motion. Our total time span covers more than one
cycle, and the orbital elements are given in Table~\ref{tab:Orbital}. 
The nebular velocities have a stable radial velocity with a 
standard deviation of 350 m\,s$^{-1}$ and a slight indication of some
modulation in antiphase.
The errors on the orbital elements were computed using a Monte Carlo (MC) error simulation in which we
used the standard deviation of the residuals after orbit removal as
a proxy of the $\sigma$ value in our simulation.  Our $\chi^2$
  values yield eccentricities up to around e = 0.19 of the different equivalent
MC datasets, but because of the
  scatter and the limited number of cycles covered, the eccentricity
  is only poorly determined. A possible eccentric fit is not significant
  according to the classical statistical test
  \citep{lucy71}. 

\begin{figure}
\begin{center}
\resizebox{.9\hsize}{!}{ \includegraphics{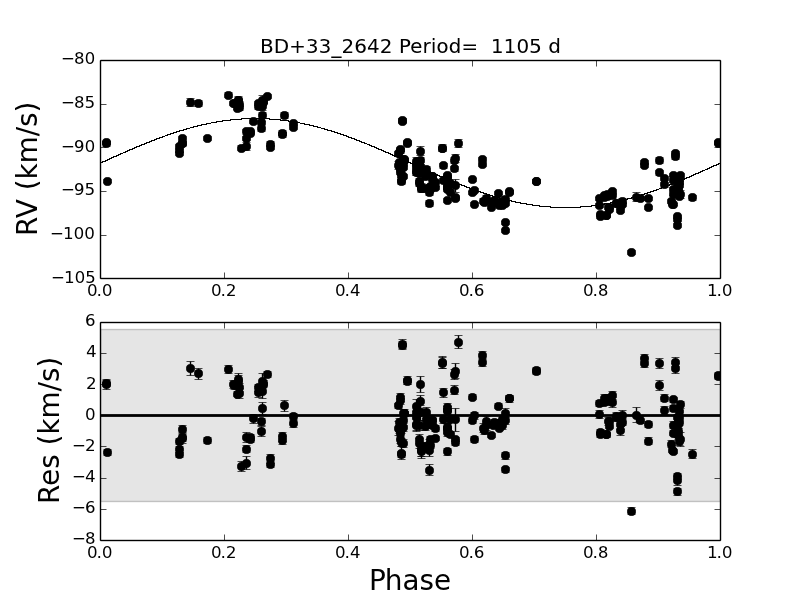}}
\caption{ Top panel: barycentric radial-velocity measurements of \bd\, 
folded on the orbital period. The full line represents the Keplerian circular
  orbit. The bottom panel shows the residuals, and the grey area is the
  three-sigma level of the residuals.
}
\label{fig:BDPhase}
\end{center}
\end{figure}

\begin{table}
\caption{Orbital elements of \bd. }
\centering
\label{tab:Orbital}                    
\begin{tabular}{lrr}        
\hline\hline                 
                        &       & $\sigma$  \\ \hline
Period (d)              & 1105. & 24.  \\ 
e                       & 0     & imposed  \\ 
K (km s$^{-1}$)         & 5.10   & 0.25  \\
$\gamma$ (km s$^{-1}$)  & -91.86 & 0.15 \\
JD(0)                   & 2456524.0 & inferior conjuction  \\
a\,$\sin{\rm i}$ (A.U.) & 0.52  & 0.02 \\
f(M) (M$_{\odot}$)     & 0.015 & 0.002 \\ 
RMS residuals  ( km\,s$^{-1}$)        & 1.8  & \\ \hline\hline
\end{tabular}    
\end{table}

The fit that imposes a circular orbit with a period of 1105 days has a fractional variance reduction of 74 \%
and the residuals have a standard deviation of 1.8 km\,s$^{-1}$, clearly
larger than the intrinsic uncertainty of the velocity determinations. 
The shorter term variability, probably of atmospheric origin, was also
reported by \cite{demarco07}.

When we add the radial velocities obtained by \cite{demarco07} to
increase the total monitoring length by some 1500 days, we obtain very
similar orbital elements with a slight increase of the most likely
orbital period to 1136 days.  The disadvantage is that these data come
from another instrument and the velocities were obtained with a
different method. A potential systematic difference between the two
instruments was not taken into account. What \cite{demarco04,
  demarco07} observed as radial-velocity offsets between the mean
velocities of their different runs is caused by orbital motion.  The
phased radial-velocity curve of the HERMES data alone is given in
Fig.~\ref{fig:BDPhase}.  With the phase dispersion minimisation
\citep{stellingwerf78} technique, we were unable to find a significant
frequency in these residuals. When we assume that the primary has an
average mass of a WD (0.6 M$_{\odot}$), the mass function of 0.015
$\pm$ 0.002 M$_{\odot}$ translates into a companion mass of 0.22
M$_{\odot}$ ($i$ = 90$^{\circ}$) and 0.26 M$_{\odot}$, for 
an inclination of 60$^{\circ}$.

\section{Analysis of HD\,112313} \label{sect:LoTr5}

\begin{figure}
\begin{center}
\resizebox{0.70\hsize}{!}{ \includegraphics{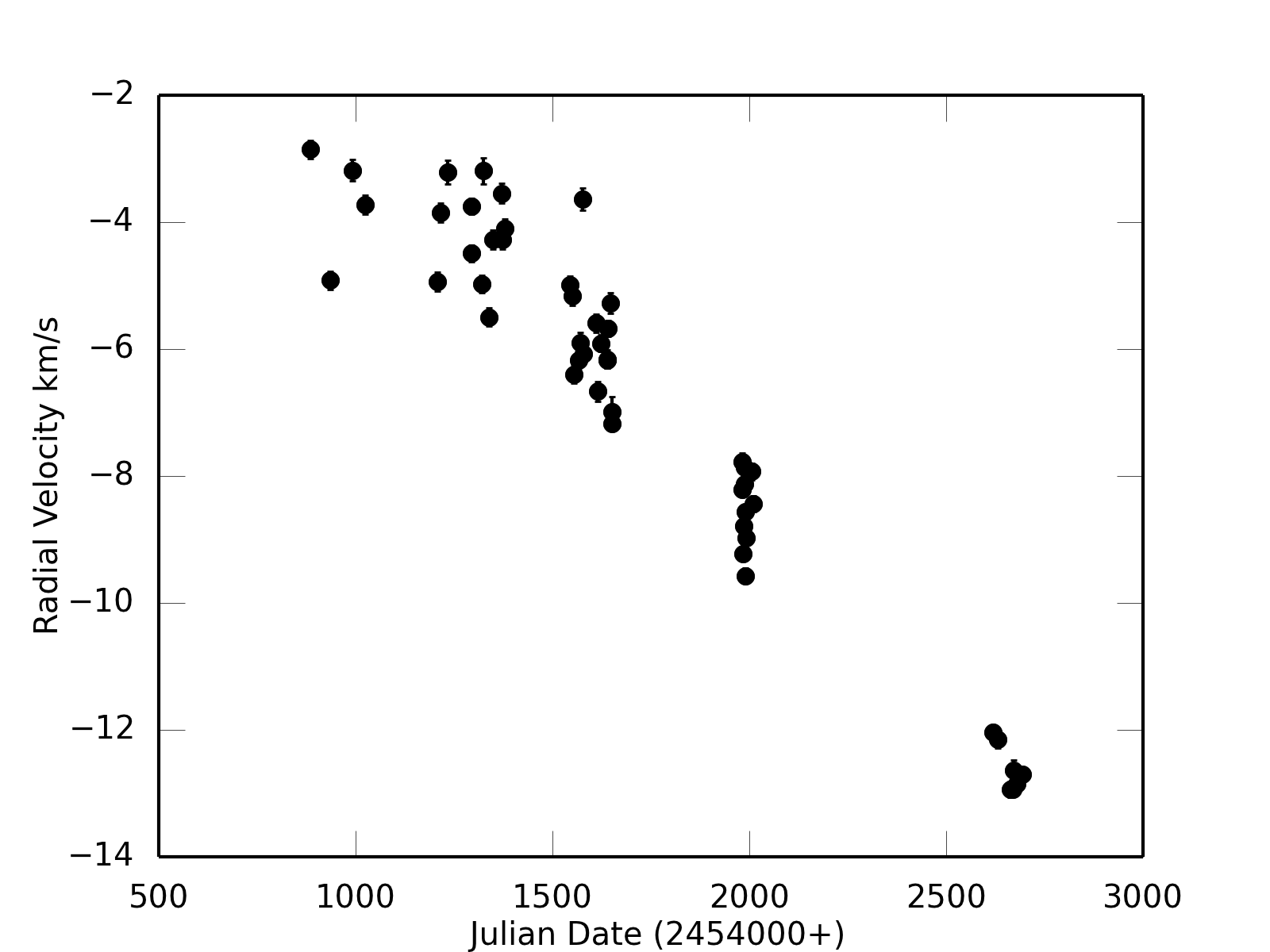}}
\caption{Time series of radial-velocity measurements of HD\,112313.}
\label{fig:LoTr5Time}
\end{center}
\end{figure}

Similar to what we did for \bd, we optimised a specific spectral
mask for HD\,112313. Starting from a well-exposed spectrum, we selected 135
relatively isolated lines in the region between 4750 and 6550\AA. The
dominant component is a rapidly rotating star, and the
cross-correlation function was fitted with a rotation profile. In
Fig.~\ref{fig:LoTr5Time}, we show the time series of our
measurements. The long-term trend is clearly visible, but despite our
monitoring efforts that were as long as 1807\,days, we have not yet covered a full
cycle. The peak-to-peak amplitude is about 10 km\,s$^{-1}$. The
$v\sin\,i$\, of 66.2 $\pm$ 0.7 km\,s$^{-1}$ agrees with the value in the
literature \citep{strassmeier97}.

We can estimate a mass function of 0.05 M$_{\odot}$\,
by assuming an amplitude of 5 km\,s$^{-1}$ and a period that is twice
as long as the total time we covered till now. The amplitude is a lower limit, and
a probable increase impacts strongly on the mass function with its
dependency on the third power. The dependency on a different orbital
period is much weaker.

Assuming that the orbital plane is co-planar with the waist of the
bipolar nebula \citep{graham04}, our inclination angle to the system is
only 17$^{\circ}$. With a typical mass of 1.1 M$_{\odot}$ for the G star and
the above mass function, the mass of the hot source in the system must
be 3.5 M$_{\odot}$, clearly incompatible with a WD mass. A more
massive G giant in the system would increase this mass even more.
The most likely larger mass function will exacerbate this conflict as well.
We conclude that either the inclination angle of the bipolar
model of \cite{graham04} is incorrect
or that the orbital plane is not co-planar with the waist of the
bipolar nebula. The alternative is that we see a triple system in which the
hot component is a double star with a combined mass of 3.5
M$_{\odot}$. The close companion of the hot WD must then be an object
of some two to three solar masses.

Obviously, we should continue our efforts to monitor this object 
to cover a full orbital cycle, which will allow us to constrain the dynamics of
the system better.

\section{Results and discussion} \label{sect:results}

We reported here on the detection of two wide spectroscopic binaries
among the central stars of PNe: one with a period of about 1100 days
and one whose orbital period is longer than 1800 days. To our
knowledge these are the first spectroscopic binaries among central
stars of PNe with orbits longer than several weeks. Radial-velocity
monitoring nicely complements the photometric techniques, which are
only sensitive to systems that spiraled-in during a CE phase
\citep[e.g.][and references therein]{miszalski12}.

The separations are puzzling because any binary with a separation
within approximately several giant radii will be tidally captured and,
if unstable mass transfer ensues, a common envelope arises with a
dramatic shortening of the orbital separation, depending also on the
envelope binding energy. The separation indicates that either the
common envelope was rapidly ejected without much in-spiralling, or
that a form of stable mass transfer occurred.  Population-synthesis
models \citep[e.g.][and references therein for an overview]{keller14}
predict a CE channel to be representative of only 14-15\% of all PNe,
which includes a small fraction that are formed through  post-RGB interaction
\citep{hall13}.  The binary channel for PNe formation also predicts
wider systems that primarily interacted by wind accretion
\citep[e.g.][]{soker97}. The final period distribution of binary PNe
is predicted to be bi-modal, in which the CE channel populates the
short-period binaries and the wider systems interacted by wind
interaction. The orbital periods that are least often 
predicted are in the
middle of the bi-modal distribution and occur at about
1000 days \citep{keller14}.  Wide binaries are also expected to exist
among central stars of PNe, but the periods found by us here fall in
the minimum of the predicted period distribution. Obviously, two
systems are not enough to draw far-reaching conclusions.

It is a long-standing problem that the predictions of
population-synthesis computation also starkly contrast with the periods
found in post-AGB binaries \citep[e.g.][]{vanwinckel09} and those
found in wide binaries with WD companions such as Ba stars
\citep[e.g.][]{jorissen99}. The periods (and eccentricities) found in
these systems are not predicted by the population-synthesis models because
they almost all fall in the minimum of the expected period
distribution.  Period-wise the PNe binaries described here form a
natural link to the post-AGB binaries (as potential precursors) and
Ba-stars (as potential successors). The dynamic information from the
orbits lead to interesting conclusions:

\begin{itemize}
\item
\bd\, is a spectrophotometric standard star and one of the few PNe in the
Galactic halo. It is a binary
with abundance anomalies, very similar to what is found in 
many Galactic post-AGB binaries with
circumbinary discs. The orbit of \bd\, is similar to the orbits found
in the post-AGB binaries. This object is therefore a natural progeny of
the class of depleted F-K type post-AGB  binaries with circumstellar
discs \citep{vanwinckel03} and
shows that the latter can also become PNe.

\item
Although the full orbit of HD\,112313 is not yet covered, the
lower limit of the mass function led to a conflict between the
nebular and orbital properties. The obtained mass function
implies that either the orbital plane is not co-planar with the waist
of the bipolar nebula, or that the WD is member of a close binary
with a total mass of 3.5 M$_{\odot}$. The cool Ba-rich
component of the system must be spun-up by wind accretion, and we
disproved the statements in the literature that the rapidly rotating G
star may be a close binary. 
\end{itemize}

Our project shows that detecting radial-velocity orbits in central
stars of PNe is possible but intensive
long-term monitoring is clearly needed. This requires a dedicated
long-term observational project. Many prime candidates for PNe with
binary central stars can be found, but they are very faint
\citep[e.g.][]{demarco13} and will require regular telescope time on
large infrastructure.

\begin{acknowledgements}
  The authors thank the referee Orsola De Marco for the careful 
  reading of the manuscript, the suggestions for improvement and the very quick referee report.
  Based on observations obtained with the HERMES spectrograph, which
  is supported by the Fund for Scientific Research of Flanders (FWO),
  Belgium , the Research Council of KU Leuven, Belgium, the Fonds
  National de la Recherche Scientifique (FNRS), Belgium, the Royal
  Observatory of Belgium, the Observatoire de Gen\`eve, Switzerland
  and the Th\"uringer Landessternwarte Tautenburg, Germany.  The
  Mercator telescope is operated thanks to grant number G.0C31.13 of
  the FWO under the 'Big Science' initiative of the Flemish
  governement.  HVW acknowledges support from The Research Council of
  the KU Leuven under grant number GOA/2013/012.  The authors want to
  thank all observers of the HERMES consortium institutes (KU Leuven,
  ULB, Royal Observatory (Belgium) and Sternwarte Tautenburg
  (Germany)), who contributed to this monitoring programme.

\end{acknowledgements}

\bibliographystyle{aa}

\end{document}